\documentclass[aps,prl,twocolumn, superscriptaddress,showpacs]{revtex4-2}

\usepackage{xcolor} 
\usepackage{graphicx} 
\usepackage{dcolumn}  
\usepackage[caption=false]{subfig}
\graphicspath{{ps}}

\begin{document}

\title{\quad\\[1.0cm]Search for the decay $B^{0}\rightarrow K^{\ast 0}\tau^{+}\tau^{-}$ at the Belle experiment}

\noaffiliation
\affiliation{Department of Physics, University of the Basque Country UPV/EHU, 48080 Bilbao}
\affiliation{University of Bonn, 53115 Bonn}
\affiliation{Brookhaven National Laboratory, Upton, New York 11973}
\affiliation{Budker Institute of Nuclear Physics SB RAS, Novosibirsk 630090}
\affiliation{Faculty of Mathematics and Physics, Charles University, 121 16 Prague}
\affiliation{Chonnam National University, Gwangju 61186}
\affiliation{University of Cincinnati, Cincinnati, Ohio 45221}
\affiliation{Deutsches Elektronen--Synchrotron, 22607 Hamburg}
\affiliation{Department of Physics, Fu Jen Catholic University, Taipei 24205}
\affiliation{Key Laboratory of Nuclear Physics and Ion-beam Application (MOE) and Institute of Modern Physics, Fudan University, Shanghai 200443}
\affiliation{Gifu University, Gifu 501-1193}
\affiliation{SOKENDAI (The Graduate University for Advanced Studies), Hayama 240-0193}
\affiliation{Gyeongsang National University, Jinju 52828}
\affiliation{Department of Physics and Institute of Natural Sciences, Hanyang University, Seoul 04763}
\affiliation{University of Hawaii, Honolulu, Hawaii 96822}
\affiliation{High Energy Accelerator Research Organization (KEK), Tsukuba 305-0801}
\affiliation{J-PARC Branch, KEK Theory Center, High Energy Accelerator Research Organization (KEK), Tsukuba 305-0801}
\affiliation{National Research University Higher School of Economics, Moscow 101000}
\affiliation{Forschungszentrum J\"{u}lich, 52425 J\"{u}lich}
\affiliation{IKERBASQUE, Basque Foundation for Science, 48013 Bilbao}
\affiliation{Indian Institute of Science Education and Research Mohali, SAS Nagar, 140306}
\affiliation{Indian Institute of Technology Bhubaneswar, Satya Nagar 751007}
\affiliation{Indian Institute of Technology Guwahati, Assam 781039}
\affiliation{Indian Institute of Technology Hyderabad, Telangana 502285}
\affiliation{Indian Institute of Technology Madras, Chennai 600036}
\affiliation{Indiana University, Bloomington, Indiana 47408}
\affiliation{Institute of High Energy Physics, Chinese Academy of Sciences, Beijing 100049}
\affiliation{Institute of High Energy Physics, Vienna 1050}
\affiliation{Institute for High Energy Physics, Protvino 142281}
\affiliation{INFN - Sezione di Napoli, I-80126 Napoli}
\affiliation{INFN - Sezione di Roma Tre, I-00146 Roma}
\affiliation{INFN - Sezione di Torino, I-10125 Torino}
\affiliation{Advanced Science Research Center, Japan Atomic Energy Agency, Naka 319-1195}
\affiliation{J. Stefan Institute, 1000 Ljubljana}
\affiliation{Institut f\"ur Experimentelle Teilchenphysik, Karlsruher Institut f\"ur Technologie, 76131 Karlsruhe}
\affiliation{Kavli Institute for the Physics and Mathematics of the Universe (WPI), University of Tokyo, Kashiwa 277-8583}
\affiliation{Kennesaw State University, Kennesaw, Georgia 30144}
\affiliation{Kitasato University, Sagamihara 252-0373}
\affiliation{Korea Institute of Science and Technology Information, Daejeon 34141}
\affiliation{Korea University, Seoul 02841}
\affiliation{Kyoto Sangyo University, Kyoto 603-8555}
\affiliation{Kyungpook National University, Daegu 41566}
\affiliation{Universit\'{e} Paris-Saclay, CNRS/IN2P3, IJCLab, 91405 Orsay}
\affiliation{P.N. Lebedev Physical Institute of the Russian Academy of Sciences, Moscow 119991}
\affiliation{Faculty of Mathematics and Physics, University of Ljubljana, 1000 Ljubljana}
\affiliation{Ludwig Maximilians University, 80539 Munich}
\affiliation{Luther College, Decorah, Iowa 52101}
\affiliation{Malaviya National Institute of Technology Jaipur, Jaipur 302017}
\affiliation{Faculty of Chemistry and Chemical Engineering, University of Maribor, 2000 Maribor}
\affiliation{Max-Planck-Institut f\"ur Physik, 80805 M\"unchen}
\affiliation{School of Physics, University of Melbourne, Victoria 3010}
\affiliation{University of Mississippi, University, Mississippi 38677}
\affiliation{University of Miyazaki, Miyazaki 889-2192}
\affiliation{Moscow Physical Engineering Institute, Moscow 115409}
\affiliation{Graduate School of Science, Nagoya University, Nagoya 464-8602}
\affiliation{Universit\`{a} di Napoli Federico II, I-80126 Napoli}
\affiliation{Nara Women's University, Nara 630-8506}
\affiliation{National Central University, Chung-li 32054}
\affiliation{National United University, Miao Li 36003}
\affiliation{Department of Physics, National Taiwan University, Taipei 10617}
\affiliation{H. Niewodniczanski Institute of Nuclear Physics, Krakow 31-342}
\affiliation{Nippon Dental University, Niigata 951-8580}
\affiliation{Niigata University, Niigata 950-2181}
\affiliation{Novosibirsk State University, Novosibirsk 630090}
\affiliation{Osaka City University, Osaka 558-8585}
\affiliation{Pacific Northwest National Laboratory, Richland, Washington 99352}
\affiliation{Panjab University, Chandigarh 160014}
\affiliation{Peking University, Beijing 100871}
\affiliation{University of Pittsburgh, Pittsburgh, Pennsylvania 15260}
\affiliation{Punjab Agricultural University, Ludhiana 141004}
\affiliation{Research Center for Nuclear Physics, Osaka University, Osaka 567-0047}
\affiliation{Meson Science Laboratory, Cluster for Pioneering Research, RIKEN, Saitama 351-0198}
\affiliation{Dipartimento di Matematica e Fisica, Universit\`{a} di Roma Tre, I-00146 Roma}
\affiliation{Department of Modern Physics and State Key Laboratory of Particle Detection and Electronics, University of Science and Technology of China, Hefei 230026}
\affiliation{Seoul National University, Seoul 08826}
\affiliation{Showa Pharmaceutical University, Tokyo 194-8543}
\affiliation{Soongsil University, Seoul 06978}
\affiliation{Sungkyunkwan University, Suwon 16419}
\affiliation{School of Physics, University of Sydney, New South Wales 2006}
\affiliation{Tata Institute of Fundamental Research, Mumbai 400005}
\affiliation{Department of Physics, Technische Universit\"at M\"unchen, 85748 Garching}
\affiliation{Toho University, Funabashi 274-8510}
\affiliation{Department of Physics, Tohoku University, Sendai 980-8578}
\affiliation{Earthquake Research Institute, University of Tokyo, Tokyo 113-0032}
\affiliation{Department of Physics, University of Tokyo, Tokyo 113-0033}
\affiliation{Tokyo Institute of Technology, Tokyo 152-8550}
\affiliation{Tokyo Metropolitan University, Tokyo 192-0397}
\affiliation{Virginia Polytechnic Institute and State University, Blacksburg, Virginia 24061}
\affiliation{Wayne State University, Detroit, Michigan 48202}
\affiliation{Yamagata University, Yamagata 990-8560}
\affiliation{Yonsei University, Seoul 03722}
  \author{T.~V.~Dong}\affiliation{Key Laboratory of Nuclear Physics and Ion-beam Application (MOE) and Institute of Modern Physics, Fudan University, Shanghai 200443} 
  \author{T.~Luo}\affiliation{Key Laboratory of Nuclear Physics and Ion-beam Application (MOE) and Institute of Modern Physics, Fudan University, Shanghai 200443} 
  \author{I.~Adachi}\affiliation{High Energy Accelerator Research Organization (KEK), Tsukuba 305-0801}\affiliation{SOKENDAI (The Graduate University for Advanced Studies), Hayama 240-0193} 
  \author{H.~Aihara}\affiliation{Department of Physics, University of Tokyo, Tokyo 113-0033} 
  \author{D.~M.~Asner}\affiliation{Brookhaven National Laboratory, Upton, New York 11973} 
  \author{H.~Atmacan}\affiliation{University of Cincinnati, Cincinnati, Ohio 45221} 
  \author{V.~Aulchenko}\affiliation{Budker Institute of Nuclear Physics SB RAS, Novosibirsk 630090}\affiliation{Novosibirsk State University, Novosibirsk 630090} 
  \author{T.~Aushev}\affiliation{National Research University Higher School of Economics, Moscow 101000} 
  \author{R.~Ayad}\affiliation{Department of Physics, Faculty of Science, University of Tabuk, Tabuk 71451} 
  \author{V.~Babu}\affiliation{Deutsches Elektronen--Synchrotron, 22607 Hamburg} 
% ...
\collaboration{The Belle Collaboration}

\begin{abstract}
  This letter presents a search for the rare flavor-changing neutral current process $B^{0}\rightarrow K^{\ast 0}\tau^{+}\tau^{-}$
  using data taken with the Belle detector at the KEKB asymmetric energy $e^{+}e^{-}$
  collider. The analysis is based on the entire $\Upsilon(4S)$ resonance data sample
  of 711 $\rm fb^{-1}$, corresponding to $772\times 10^{6} B \bar{B}$ pairs. In our
  search we fully reconstruct the companion $B$ meson produced in the process
  $e^{+}e^{-}\rightarrow\Upsilon(4S)\rightarrow B\bar{B}$ from
  its hadronic decay modes, and look for the decay $B^{0}\rightarrow K^{\ast 0}\tau^{+}\tau^{-}$
  in the rest of the event. No evidence for a signal is found. We report an upper
  limit on the branching fraction
  $\mathcal{B}({B^{0}\rightarrow K^{\ast 0}\tau^{+}\tau^{-}})<3.1\times 10^{-3}$
  at 90\% confidence level. This is the first direct limit
  on $\mathcal{B}({B^{0}\rightarrow K^{\ast 0}\tau^{+}\tau^{-}})$.
\end{abstract}
\maketitle
\tighten
The decay $B^{0}\rightarrow K^{\ast 0}\tau^{+}\tau^{-}$
(charge-conjugate processes are implied throughout this letter)
is of interest for the testing of Lepton Flavor Universality (LFU)
and for searches of physics beyond the Standard Model (SM).
This decay is highly suppressed in the SM and can only
proceed via a flavor-changing neutral current, with a predicted
branching fraction of order $\mathcal{O}(10^{-7})$ \cite{btosll_sm1}.
The branching fraction can be enhanced if new physics (NP) effects contribute
\cite{btosll_sm2, btosll_theory, btosll_NP, btosll_lepto}.
The flavor-changing neutral current processes such as
$B^{0}\rightarrow K^{\ast 0}\tau^{+}\tau^{-}$ can provide
very powerful tests for the SM and its extensions.
In particular, the decay is a third-generation equivalent of the
$B^{0}\rightarrow K^{\ast 0}\ell^{+}\ell^{-}$ decay, where $\ell$
is an electron or a muon. Hence, compared with electron and muon
modes, the decay is expected to be more sensitive to new physics
in a model which has a coupling proportionate to the particle mass \cite{btosll_2HDM}
or only couples to the third generation \cite{btosll_eft}.

Semileptonic $B$ decay measurements in recent years show
significant deviations from SM expectations, for both charged
and neutral current transitions. The first type of transition 
has been measured in the decay $b\rightarrow c\ell\bar{\nu}_{\ell}$ via
$R(D^{(*)})=[\mathcal{B}(B\rightarrow D^{(*)}\tau^{+}\nu_{\tau})]/[\mathcal{B}(B\rightarrow D^{(*)}\ell^{+}\nu_{\ell})]$
by the BaBar \cite{RDst_babar1, RDst_babar2}, Belle \cite{RDst_belle1, RDst_belle2, RDst_belle3, RD_belle4} and LHCb~\cite{RDst_lhcb1, RDst_lhcb2} experiments. While these decays are tree-level
processes, which are not very sensitive to NP, the measured
results show a deviation of about three standard deviations, 3$\sigma$,
from the SM predictions (combined significance) \cite{hflav2021}.
The neutral current transition $b\rightarrow s \ell^{+} \ell^{-}$
is highly suppressed in the SM and very sensitive to NP.
The LFU ratio between muon and electron in the decay mode $B\rightarrow K^{(*)}\ell^{+}\ell^{-}$
as measured by Belle \cite{RK_belle2009, RK_belle2021, Angular_PQ45_belle2017}
and BaBar \cite{RK_babar2012} are consistent with the SM, while
LHCb result \cite{RK_lhcb, RKst_lhcb, RK_lhcb2021} is 3.1$\sigma$
lower than the SM prediction. Many theoretical models are introduced to
explain these anomalies such as the NP contribution to
the Wilson coefficients \cite{btosll_theory, btosll_NP} and
the leptoquark model \cite{btosll_lepto}. These approaches lead
to an enhancement of the $b\rightarrow s\tau^{+}\tau^{-}$
branching fraction up to $1$-$5 \times 10^{-4}$, 3 orders of magnitude larger than the SM predictions.
The predicted branching fraction of $B^{0}\rightarrow K^{\ast 0}\tau^{+}\tau^{-}$
is larger than that of $B^{+}\rightarrow K^{+}\tau^{+}\tau^{-}$ as shown in Ref.~\cite{btosll_theory}.

The presence of at least two neutrinos in the final state originating from
the decays of $\tau^{+}\tau^{-}$ pair make analysis of the decay challenging.
To date only a search for the decay $B^{+}\rightarrow K^{+}\tau^{+}\tau^{-}$
has been conducted by the BaBar collaboration setting an upper limit
$\mathcal{B}(B^{+}\rightarrow K^{+}\tau^{+}\tau^{-})<2.25\times 10^{-3}$
at 90\% confidence level (CL) \cite{b2ktt_babar}.

In this letter, we present the first search for the rare decay
$B^{0}\rightarrow K^{\ast 0}\tau^{+}\tau^{-}$.
Our analysis is based on the complete data set collected at the
center of mass (c.m.) energy equal to the $\Upsilon(4S)$ resonance mass
by the Belle detector \cite{belle_det} at the KEKB asymmetric-energy
$e^{+}e^{-}$ collider \cite{kekb}. This data sample corresponds to
an integrated luminosity of 711 $\rm fb^{-1}$, containing
$772\times 10^{6} B\bar{B}$ pairs. We use a full reconstruction technique
\cite{fullrecon} in this analysis where the companion
$B$ meson in the process $e^{+}e^{-}\rightarrow\Upsilon(4S)\rightarrow B\bar{B}$
is reconstructed in hadronic decay modes, referred to
as $B_{\rm tag}$. We then search for the signal $B$ meson, $B_{\rm sig}$,
in the rest of the event not used in the $B_{\rm tag}$ reconstruction.

The Belle detector~\cite{belle_det} is a large-solid-angle magnetic
spectrometer consisting of a silicon vertex detector (SVD),
a 50-layer central drift chamber (CDC), an array of
aerogel threshold Cherenkov counters (ACC),
a barrel-like arrangement of time-of-flight
scintillation counters (TOF), and an electromagnetic calorimeter
comprised of CsI(Tl) crystals (ECL). All these components are
located inside a superconducting solenoid coil that provides a 1.5~T
magnetic field.  An iron flux-return located outside of the coil
is instrumented with resistive plate chambers to detect $K_L^0$ mesons and to identify
muons (KLM).

We use Monte Carlo (MC) simulation samples, generated with EvtGen~\cite{evtgen},
to optimize the signal selection, determine the selection efficiencies,
as well as to obtain the signal and background fitting models. The detector response
is simulated using GEANT3~\cite{geant3}. Simulated events are overlaid
with random trigger data taken for each run period to reproduce the effect
of beam-associated backgrounds. A signal sample containing 50 million 
$\Upsilon(4S)\rightarrow B^{0}\bar{B}^{0}$ events is generated
where one $B$ decays to all possible final states, according to its measured or
estimated branching fractions~\cite{pdg2020}, and the other decays via
$B^{0}\rightarrow K^{\ast 0}\tau^{+}\tau^{-}$, using the model described
in Ref.~\cite{BTOSLLBALL1}. Background MC samples consist of $B^{+}B^{-}$, $B^{0}\bar{B}^{0}$,
and continuum $e^{+}e^{-}\rightarrow q\bar{q}$ ($q= u, d, s, c$),
where the size of each sample is six times larger than that of collision data.
Rare $B$ meson decay processes such as charmless hadronic, radiative,
and electroweak decays are simulated separately in a sample designated
{\it Rare B}. Semileptonic $b\rightarrow u\ell\nu$ decays are simulated
in a dedicated $u\ell\nu$ sample. The sizes of the $Rare~B$ and $u\ell\nu$
samples are 50 and 20 times larger than that of collision data, respectively.

A candidate $B_{\rm tag}$ meson is reconstructed in one of the 489 hadronic
decay channel using a hierarchical NeuroBayes-based (NB) full-reconstruction
algorithm \cite{fullrecon}. In this algorithm, the continuum backgrounds are suppressed
by employed event shape variables such as the polar angle of $B_{\rm tag}$, 
the cosine of the angle between the thrust axis \cite{thrust} and z-direction,
and the modified second Fox-Wolfram moment \cite{sfw}.
All the input variables which used during the reconstruction are mapped to a
single classifier output, ${\cal O}_{\rm NB}$, which represents the
quality of $B_{\rm tag}$, ranges from zero for combinatorial background
and continuum events to unity for an unambiguous $B_{\rm tag}$.
Event selection also exploits the energy difference
$\Delta E = E_{\rm Btag} - E_{\rm cm}/2$ and the beam-energy-constrained mass
$M_{\rm bc} =\sqrt{(E_{\rm cm}/2)^2/c^{4}-|\vec{p}_{\rm Btag}|^{2}/c^{2}}$,
where $E_{\rm cm}$ is the $e^{+}e^{-}$ energy, and $E_{\rm Btag}$ and
$\vec{p}_{\rm Btag}$ are the reconstructed energy and momentum of the $B_{\rm tag}$
candidate, respectively. All the quantities are measured in the c.m.\ frame.
We require each $B_{\rm tag}$ candidate to satisfy
$\ln({\cal O}_{\rm NB})>-7$, $|\Delta E|<0.06$ GeV, and
$5.275 <M_{\rm bc}<5.290$ GeV/$c^{2}$. The net tagging efficiency which is
defined as number of truly reconstructed B-tag divided for total number of generated event is 0.24\%.
It is slightly higher than that reported in Ref.~\cite{fullrecon}, due to lower
average particle multiplicity in this signal sample compared to generic sample.
The signal side of the $B^{0}\rightarrow K^{\ast 0}\tau^{-}\tau^{+}$ sample
contains of leptons and only two hadron tracks.
The possibility of the interference from signal side to the tag side
reconstruction is lower than that of the generic samples, where both $B$
mesons decay generically. The tagging efficiency is calibrated by comparing
the known branching fraction from PDG \cite{pdg2020} of the decays
$B\rightarrow D^{(*)}\ell\nu_{\ell}$ and the measured values
which use this hadron tag reconstruction method~\cite{tag_eff_corr}.

For events where a $B_{\rm tag}$ is reconstructed, we search for the decay
$B^{0}\rightarrow K^{\ast 0}\tau^{+}\tau^{-}$ in the rest of the event.
The remaining tracks are examined to remove duplicate ones
due to the curling of low transverse momentum particles ($p_{t}<0.3$ GeV/$c$).
A pair of tracks is considered as duplicate if the cosine of the angle between them
is either larger than 0.9 or smaller than 0.1, and the difference in transverse
momentum is less than 0.1 GeV/$c$. All tracks are constrained to originate
from the interaction point (IP) by the requirements $|dr|<2$ cm and $|dz|<4$ cm,
where $dr$ and $dz$ are the impact parameter with respect to IP
in the transverse and longitudinal directions, respectively.
We select events as signal candidates if there are
four remaining tracks with zero net charge. The number of signal candidates doubles
after removing duplicate tracks.

We reconstruct candidate $K^{\ast0}$ mesons from $K^{\ast 0}\rightarrow K^{+}\pi^{-}$ decays
using two of the four remaining tracks. We identify kaons and pions
based on combined information from the CDC, ACC, and TOF \cite{belle_pid}.
A charged track is identified as a kaon if the likelihood ratio
$\mathcal{R}_{K}=\mathcal{L}_{K}/(\mathcal{L}_{K}+\mathcal{L}_{\pi})>0.6$,
and as a pion if $\mathcal{R}_{K} < 0.4$,
where $\mathcal{L}_{i}$ is the PID likelihood for the particle type $i$.
The momenta of $K^{+}$ and $\pi^{-}$ candidates are required to be greater than 0.1 GeV/$c$.
The flavor of the reconstructed $K^{\ast 0}$ and hence the corresponding flavor of
$B_{\rm sig}$ is required to be opposite to that of $B_{\rm tag}$.
This requirement rejects 20\% of the events.
We fit the vertex for $K^{\ast 0}$ candidates,
and reject candidates if the vertex fit fails. If more than one
$K^{\ast 0}$ candidate is successfully reconstructed, the one having the reconstructed mass
closest to the known $K^{\ast 0}$ mass is retained. We require the mass of the reconstructed 
$K^{\ast 0}$ candidate to be in the range [0.8, 1.0] GeV/$c^{2}$,
which is approximately twice the decay width of $K^{\ast 0}$. 
We consider three $\tau$ decay modes in this analysis:
$\tau^{-}\rightarrow e^{-}\bar{\nu}_{e}\nu_{\tau}$,
$\tau^{-}\rightarrow\mu^{-}\bar{\nu}_{\mu}\nu_{\tau}$, and
$\tau^{-}\rightarrow\pi^{-}\nu_{\tau}$, resulting in six different decay topologies:
$K^{\ast 0}e^{+}e^{-}$, $K^{\ast 0}e^{\mp}\mu^{\pm}$, $K^{\ast 0}\mu^{+}\mu^{-}$,
$K^{\ast 0}e^{\mp}\pi^{\pm}$, $K^{\ast 0}\mu^{\mp}\pi^{\pm}$, and $K^{\ast 0}\pi^{+}\pi^{-}$.
We regard the two remaining tracks not used in the $B_{\rm tag}$
or the $K^{\ast 0}$ candidates as $\tau$ decay products. The reconstructed mass
of these two tracks is required to be less than 2.5 GeV/$c^{2}$.

All the tracks and clusters in a signal event are used for the reconstruction
of $B_{\rm tag}$ and $B_{\rm sig}$. However, there are still tracks and clusters 
from beam background and possible duplicate tracking reconstruction. We require
that there be no extra $\pi^{0}$ nor $K_{S}^{0}$ candidates, and at most one
$K_{L}^{0}$ candidate cluster, to allow for beam-associated backgrounds
or electronic noise. We reconstruct $K_{L}^{0}$ candidates
based on the hit patterns in the KLM subdetector not associated
with any charged track. A $\pi^{0}$ candidate is reconstructed from
$\pi^{0}\rightarrow \gamma\gamma$ in which neither daughter photon is
included in the reconstructed $B_{\rm tag}$ and whose reconstructed mass
is within 25 MeV/$c^{2}$ of the nominal $\pi^{0}$
mass~\cite{pdg2020}, corresponding to $3\sigma$ of the $\pi^{0}$ mass resolution.
Energy of photon candidates must exceed 50 MeV
and we require their shower shape, characterized as the ratio of total energy detected
in a $3\times3$ versus $5\times5$ array of ECL crystals in which the center
crystal has the maximum detected energy, to be larger than 0.75.
We reconstruct candidate $K_{S}^{0}$ from $K_{S}^{0}\rightarrow \pi^{+}\pi^{-}$ decays,
where the reconstructed mass is within 15 MeV/$c^{2}$ of
the nominal mass, corresponding to $3\sigma$ of $K_{S}^{0}$ mass resolution.

We determine the number of signal candidates by fitting the distribution of extra
calorimeter energy, $E_{\rm ECL}^{\rm extra}$, which is defined as the total energy
of the neutral clusters detected in the ECL not associated with either $B_{\rm tag}$
or $B_{\rm sig}$. We reduce the contribution of
beam-associated backgrounds while estimating $E_{\rm ECL}^{\rm extra}$ by only
counting clusters with energy greater than 0.15, 0.05 and 0.10 GeV
for the backward, barrel, and forward regions, respectively.
In signal events $E_{\rm ECL}^{\rm extra}$ should be zero or have a small value due
to the residual energy from beam-associated backgrounds or mismatched tracks.
Background events tend to have larger values due to contributions
from additional neutral clusters. We select events with $E_{\rm ECL}^{\rm extra}< 0.2$ GeV
as the signal region and those with  $E_{\rm ECL}^{\rm extra}< 2$ GeV for sideband studies.
The selection criteria in this study are chosen to maximize the search sensitivity
in the signal region following the Punzi figure of merit \cite{fom_punzi}.

In the c.m.\ frame, the $B_{\rm tag}$ and $B_{\rm sig}$ have opposite flight directions,
and the $B_{\rm tag}$ is fully reconstructed and its four-vector is determined.
The momentum of $B_{\rm sig}$ is thus derived from
the $B_{\rm tag}$ reconstruction. Its direction
is opposite the $B_{\rm tag}$ and its magnitude is calculated as
$|\vec{p}_{B\rm{sig}}| =\sqrt{(E_{\rm{cm}}/2)^{2}/c^{2}-m_{B}^{2}c^{2}}$,
where $\vec{p}_{B\rm{sig}}$ is the momentum vector of $B_{\rm sig}$,
$E_{\rm{cm}}/2$ is the beam energy measured in c.m\ frame, and 
$m_{B}$ is the nominal $B^{0}$ meson mass \cite{pdg2020}. 
We calculate the $\tau^{+}\tau^{-}$ pair invariant mass, $M_{\tau^{+}\tau^{-}}$,
by subtracting the reconstructed $K^{\ast 0}$ c.m. four-vector 
from the $B_{\rm sig}$'s giving its kinematic limits.
We require $M_{\tau^{+}\tau^{-}}$ to be greater than 3.55 GeV/$c^{2}$
to suppress combinatorial background.

After the selections above, the remaining background is final-state dependent.
We classify the remaining events into signal modes based on
final-state particles for further background suppression.
We identify electron candidates using an electron likelihood ratio,
$\mathcal{R}_{e}=\mathcal{L}_{e}/(\mathcal{L}_{e}+\mathcal{L}_{\bar{e}})$,
$\bar{e}$ indicates non-electron hypothesis. $\mathcal{L}_{e}$ ($\mathcal{L}_{\bar{e}}$)
are calculated based on $dE/dx$ information from the CDC, the ratio of the
energy deposited in the ECL to the momentum measured by the CDC and SVD,
the shower shape in the ECL, hit information from the ACC, and matching
between the position of the charged track and the ECL cluster~\cite{eid}.
Muon candidates are identified using a muon likelihood ratio,
$\mathcal{R}_{\mu} = \mathcal{L}_{\mu}/(\mathcal{L}_{\mu}+\mathcal{L}_{\pi}+\mathcal{L}_{K})$, which is estimated based on the difference between the range of the
track in KLM, estimated assuming no hadronic interactions, and the actual
range observed in the KLM. A $\chi^{2}$ from
extrapolating a track to the signals identified in the KLM using
a Kalman filter also contributes to the likelihood \cite{muid}. 
Tracks are identified as electrons if $\mathcal{R}_{e}>0.8$,
as muons if not satisfying the electron requirement and have $\mathcal{R}_{\mu}>0.8$, 
and as a pion if not either an electron or a muon.
The average of electron (muon) identification efficiency for the selection
$\mathcal{R}_{e(\mu)}>0.8$ is 92 (92)\% with pion fake rate of 0.25 (2.5)\%.
In the signal decay modes $K^{\ast 0}\pi^{+}\pi^{-}$ and
$K^{\ast 0}\ell^{\pm}\pi^{\mp}$, there remains a large background contribution from
the decay $B^{0} \rightarrow D^{(*)-}\ell^{+}\nu_{\ell}$, where
$D^{(*)-}\rightarrow K^{\ast 0}\pi^{-}(\pi^{0})$.
We suppress this by requiring the invariant mass
$M_{K^{\ast 0}\pi^{-}}$ to lie outside the $D^{-}$ mass region, 
$M_{K^{\ast 0}\pi^{-}} <1.84$ GeV/$c^{2}$ or $M_{K^{\ast 0}\pi^{-}}>1.94$ GeV/$c^{2}$,
where $M_{K^{\ast 0}\pi^{-}}$ is the combination of the $K^{\ast 0}$ candidate
and a track that is opposite to the charge of the kaon candidate in the
$K^{\ast 0}$ decay. Combinatorial background is also significant
in these signal modes, and so the  ${\cal O}_{\rm NB}$ selection criterion
is tightened to ${\ln(\cal O}_{\rm NB})>-4$ for these modes.

After we apply above selection criteria, our simulation predicts that
the remaining backgrounds with low $E_{\rm ECL}^{\rm extra}$ are primarily
$B^{0}\bar{B}^{0}$ events in which a $B_{\rm tag}$ is properly reconstructed opposite
$B^{0} \rightarrow D^{-}\ell^{+}\nu_{\ell}$ decaying to
$D^{-}\rightarrow K^{\ast 0}\ell^{-}\bar{\nu}_{\ell}$.
Such events have the same final-state particles as signal events.
The different number of missing neutrinos results in a different
missing mass distribution, $M_{\rm miss}$. We calculate this
by subtracting the measured part of the four-momentum of $B_{\rm sig}$
from the derived four-momentum of $B_{\rm sig}$ from the
recoil against $B_{\rm tag}$. In addition to the
missing mass, we find $M_{K^{\ast 0}\pi^{-}}$ is also powerful 
distinguishing signal from the remaining background. For $K^{\ast 0}\ell^{-}\ell^{+}$ modes,
we calculate $M_{K^{\ast 0}\pi^{-}}$ by combining the negatively charged
lepton with the $K^{\ast 0}$ assuming a pion mass.
We optimize selection criteria based on $M_{\rm miss}^{2}$ and $M_{K^{\ast 0}\pi^{-}}$
together mode-by-mode. These are summarized in Table \ref{tab:cut}.
Since the number of missing neutrinos from the $K^{\ast 0}\pi^{+}\pi^{-}$ mode
is the same as that from the $B^{0} \rightarrow D^{-}\ell^{+}\nu_{\ell}$ background,
the $M_{\rm miss}^{2}$ is ineffective in rejecting this background.
We only apply the selection $M_{\rm miss}^{2}<9$ $\rm{GeV}^{2}/c^{4}$ to
reject combinatorial background which is significant in this mode.
Despite the continuum suppression performed by
the full reconstruction algorithm, a small fraction of continuum
events remains in the $K^{\ast 0}\pi^{+}\pi^{-}$ signal mode.  In this case,
further constraints on the event shape are imposed. Specifically,
the event thrust is required to be smaller than 0.85, the cosine of the angle
between the thrust of $B_{\rm sig}$ and that of $B_{\rm tag}$ must be
smaller than 0.85, and the modified second Fox-Wolfram moment is
required to be less than 0.4.

\begin{table}[ht]
  \caption{Summary of the selection criteria imposed on $M_{K^{\ast 0}\pi^{-}}$
    and  $M_{\rm miss}^{2}$ for each of the signal modes.}
    \begin{tabular}{ccc}  
      \hline \hline
      Signal Mode         & $M_{K^{\ast 0}\pi^{-}}$  & $M_{\rm miss}^{2}$\\
                          &  (GeV/$c^{2}$)        &   ($\rm{GeV}^{2}/c^{4}$)\\
      \hline
      $K^{\ast 0}e^{+}e^{-}$     & $>1.4$ &$>3.2$\\
      $K^{\ast 0}e^{\mp}\mu^{\pm}$   & $>1.4$ &$>1.6$\\
      $K^{\ast 0}\mu^{+}\mu^{-}$ & $>1.6$ &$>1.6$\\
      $K^{\ast 0}\pi^{\mp} e^{\pm}$  & $>1.4$ &$>2.0$\\
      $K^{\ast 0}\pi^{\mp}\mu^{\pm}$ & $>1.4$ &$>2.0$\\
      $K^{\ast 0}\pi^{+}\pi^{-}$ & $>1.5$ &$<9$\\
      \hline \hline
    \end{tabular}
    \label{tab:cut}
\end{table}

\begin{figure}[t!]
  \subfloat{%
    \includegraphics[width=9cm]{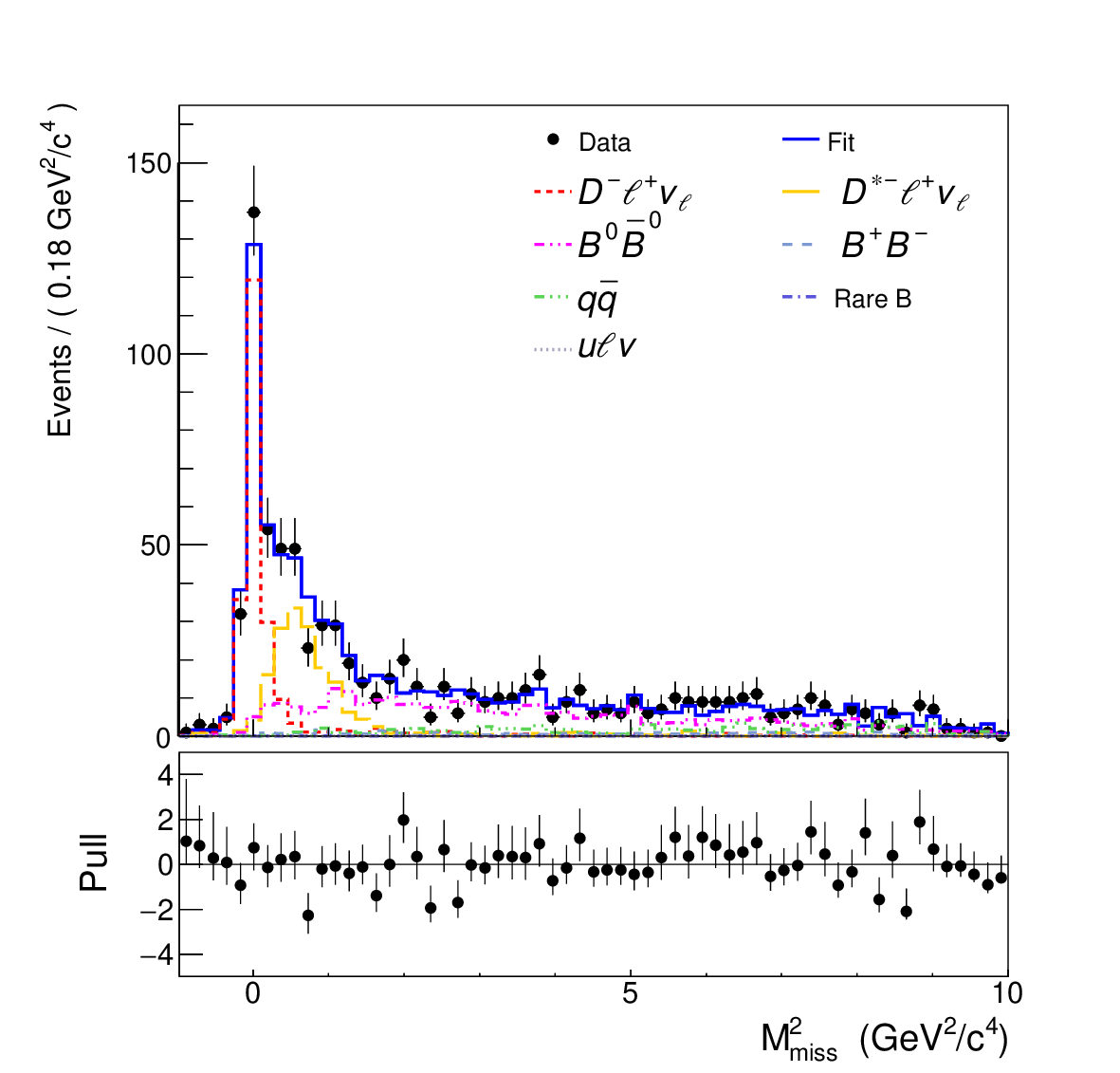}%
  }\\
  \subfloat{%
    \includegraphics[width=9cm]{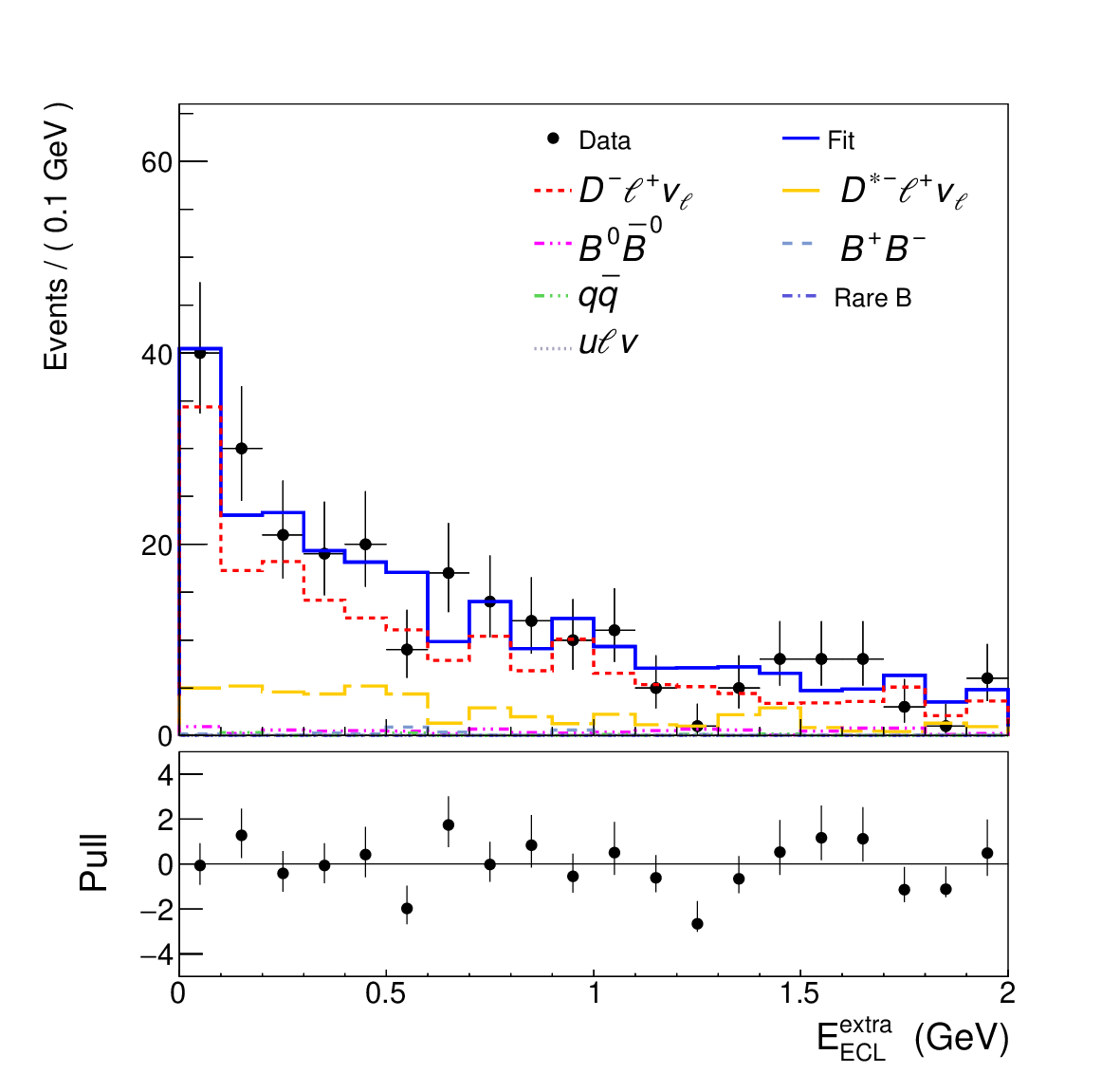}%
  }
  \caption {Fit results for $M_{\rm miss}^{2}$ (upper) and $E_{\rm ECL}^{\rm extra}$ (lower) 
    for the decays  $B^{0} \rightarrow D^{(*)}\ell\nu_{\ell}$.
    The dots with error bars represent the data, and the blue line indicates
    the fitted results. The dashed lines indicate different fit components.
    $E_{\rm ECL}^{\rm extra}$ is plotted with the selection
    $M_{\rm miss}^{2}<0.5 ~\rm{GeV}^{2}$/$c^{4}$.}
  \label{fig:fit_data_cc}
\end{figure}

We estimate the signal reconstruction efficiency after applying all of the selection criteria
and $B_{\rm tag}$ efficiency corrections. The overall selection efficiency,
determined using simulated $B^{0}\rightarrow K^{\ast 0}\tau^{+}\tau^{-}$ decays,
is approximately $(1.23\pm 0.05)\times 10^{-5}$, where the uncertainty is statistical.
The signal yield is extracted with a binned extended maximum-likelihood fit to the $E_{\rm ECL}^{\rm extra}$
distribution, with a bin width of 0.1 GeV. The probability density functions (PDFs)
for signal and background components are taken from MC expectations
after applying the B-tag efficiency correction. To reduce the uncertainty due to low statistics,
a simulation sample three times larger than the data is used to construct the background PDFs,
the signal PDF is derived from  50 million $\Upsilon(4S)\rightarrow B^{0}\bar{B}^{0}$ signal events,
and signal modes are combined in the fit. The $B^{+}B^{-}$ and $B^{0}\bar{B}^{0}$
samples are normalized to the data and their ratio is fixed in the fit.
Contributions from $Rare~B$ and $u\ell\nu_{\ell}$ components in the final sample are
negligible, and are normalized to the number of $B\bar{B}$ pairs and fixed in the fit.
We float the $B\bar{B}$, continuum, and signal normalizations.
We have validated the fitting procedure in tests with MC samples.

We test the analysis procedure and shape of the simulated $E_{\rm ECL}^{\rm extra}$
distribution using $B^{0} \rightarrow D^{-}\ell^{+}\nu_{\ell}$ decays,
with $D^{-}\rightarrow K^{\ast 0}\pi^{-}$. The analysis steps and selection
criteria for the decay are the same as those for the
$B^{0}\rightarrow K^{\ast 0}\tau^{+}\tau^{-}$ decay, except the requirement  
on $M_{\rm miss}^{2}$ is removed and the selection on $M_{K^{\ast 0}\pi^{-}}$
is reversed, requiring $1.84<M_{K^{\ast 0}\pi^{-}}< 1.94$ GeV/$c^{2}$.
We divided the sample into the two sub-samples, one with $M_{\rm miss}^{2}<0.5$
$\rm{GeV}^{2}/c^{4}$ and the other with $M_{\rm miss}^{2}>0.5$ $\rm{GeV}^{2}/c^{4}$.
The first sub-sample where events are mainly from
$B^{0} \rightarrow D^{-}\ell^{+}\nu_{\ell}$ is useful for
checking the signal shape. The latter containing mostly background events
is used for validate the background shape. Within statistics,
the signal and background models obtained from simulation
are in good agreement with the data and are used to model the signal
and background in the final fit. As a cross-check,
we measure the branching fraction of the decay
$B^{0} \rightarrow D^{-}\ell^{+}\nu_{\ell}$ from a
fit to the $E_{\rm ECL}^{\rm extra}$ distributions, similar to our search
for the decay $B^{0}\rightarrow K^{\ast 0}\tau^{+}\tau^{-}$, and also
to the $M_{\rm miss}^{2}$ distribution. Results of these fits
are shown in Fig. \ref{fig:fit_data_cc}. The branching fraction measured
by fitting to $E_{\rm ECL}^{\rm extra}$ for the first sub-sample is
$(2.45 \pm 0.17)\%$ and to $M_{\rm miss}^{2}$ is $(2.37\pm0.15)\%$,
where the quoted uncertainties are statistical only.
The results are in good agreement with the world average of $2.31\pm 0.10$\% \cite{pdg2020}.
We obtained a zero signal for the fit to $E_{\rm ECL}^{\rm extra}$ distribution
of the second sub-sample ($M_{\rm miss}^{2}>0.5 $ $\rm{GeV}^{2}/c^{4}$) as expected.

\begin{figure}[t!]
  \includegraphics[width=9cm]{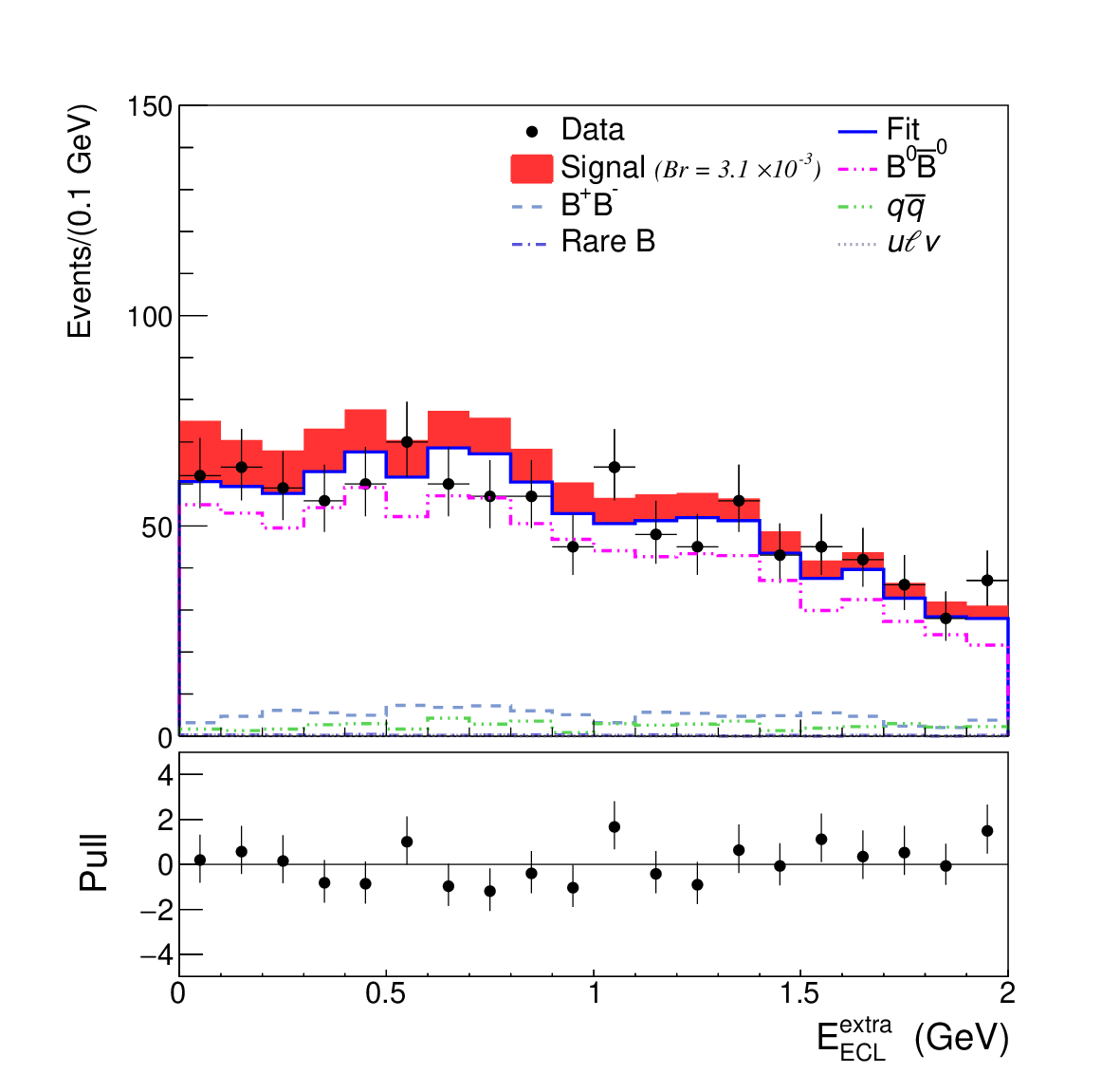}%
  \caption{Distribution of $E_{\rm ECL}^{\rm extra}$ combined for all signal modes.
    The dots with error bars show the data, the blue line shows the fitted
    results with the background-only model, and the dashed lines show fit results
    for the different components. A signal (red region) with a branching fraction
    of $3.1 \times 10^{-3}$, corresponding to the upper limit at 90\% CL, is superimposed on the top of the fit.}
  \label{fig:result}
\end{figure}

We perform the fit to $E_{\rm ECL}^{\rm extra}$ for the decay $B^{0}\rightarrow K^{\ast 0}\tau^{+}\tau^{-}$ using the procedure as described above, where all signal modes have been combined. 
The numbers of signal and background events in the signal window [0; 0.2] GeV
obtained from the fit are $N_{\rm sig}=-4.9\pm 6.0$ and $N_{\rm bkg}=122.4\pm4.9$, respectively.
We find no evidence for a signal. Data are consistent with background as shown in 
Fig.~\ref{fig:result}, where the background-only model is fitted to data and 
a signal with branching fraction of $3.1\times 10^{-3}$ is superimposed
on the top.

Systematic uncertainties on the number of background events,
the signal reconstruction efficiency, and number of $B\bar{B}$
pairs arise from several sources and affect the branching fraction
upper limit. The uncertainty on number of $B\bar{B}$ pairs is 1.8\%.
The statistical uncertainty on the selection efficiency due to
limited MC sample size is estimated to be 4.0\%. The uncertainty
associated with the $B_{\rm tag}$ efficiency is 5.1\%, which is estimated
using various decays as studied in Ref.~\cite{tag_eff_corr}. Tracking uncertainty
is assigned to be 1.4\% for the four $B_{\rm sig}$ charged tracks.
The uncertainty due to the charged track selection is estimated to be 4.1\%.
Particle identification impacts $K^{\ast 0}$ reconstruction and
signal mode separation, hence the uncertainties from electron, muon,
and pion identification are weighted following their fraction in the
signal mode. The total particle identification uncertainty is 2.55\%.
The difference in reconstruction efficiency for $\pi^{0}$
and $K_{S}^{0}$ leads to a systematic uncertainty in application of
the corresponding vetoes. Their uncertainties are estimated
to be 0.17\% and 1.56\% for $\pi^{0}$ and $K_{S}^{0}$, respectively.
The uncertainty on the branching fraction of $\tau$ is 0.57\%.
The total systematic uncertainty is 8.5\% calculated by summing
the above uncertainties in quadrature.

The systematic uncertainty due to the statistical error of the PDF
templates is estimated by varying bin contents of the templates
following the Poisson distribution and repeating the fit to the data.
This step is repeated 1000 times for each of the PDF.
The standard deviation of the number of signal distribution obtained from the fits
is considered as systematics uncertainty. The total uncertainty is 4.59 events.

The signal yield obtained from the extended maximum-likelihood fit is
translated into an upper limit on the $B\rightarrow K^{\ast 0} \tau^{+}\tau^{-}$
branching fraction using the CLs method \cite{CLs, asymptotic_cal}.
We account for statistical and systematic uncertainties on the number of background
events and signal efficiencies by modeling them as
Gaussian functions with standard deviations given by their uncertainties.
Our observed upper limit on the $B\rightarrow K^{\ast 0} \tau^{+}\tau^{-}$
branching fraction is $3.1\times 10^{-3}$ at 90\% CL.

In conclusion, we have performed a search for the decay
$B\rightarrow K^{\ast 0} \tau^{+}\tau^{-}$ using the full
Belle data set collected at the c.m.\ energy of the $\Upsilon(4S)$ resonance.
We find no signal and set an upper limit on the branching fraction
to be $3.1\times 10^{-3}$ at 90\% CL.
This is the first experimental limit on the decay
$B\rightarrow K^{\ast 0} \tau^{+}\tau^{-}$.

We thank the KEKB group for the excellent operation of the
accelerator; the KEK cryogenics group for the efficient
operation of the solenoid; and the KEK computer group, and the Pacific Northwest National
Laboratory (PNNL) Environmental Molecular Sciences Laboratory (EMSL)
computing group for strong computing support; and the National
Institute of Informatics, and Science Information NETwork 5 (SINET5) for
valuable network support.  We acknowledge support from
the Ministry of Education, Culture, Sports, Science, and
Technology (MEXT) of Japan, the Japan Society for the 
Promotion of Science (JSPS), and the Tau-Lepton Physics 
Research Center of Nagoya University; 
the Australian Research Council including grants
DP180102629, 
DP170102389, 
DP170102204, 
DP150103061, 
FT130100303; 
Austrian Federal Ministry of Education, Science and Research (FWF) and
FWF Austrian Science Fund No.~P~31361-N36;
the National Natural Science Foundation of China under Contracts
No.~11435013,  
No.~11475187,  
No.~11521505,  
No.~11575017,  
No.~11675166,  
No.~11705209;  
Key Research Program of Frontier Sciences, Chinese Academy of Sciences (CAS), Grant No.~QYZDJ-SSW-SLH011; 
the  CAS Center for Excellence in Particle Physics (CCEPP); 
the Shanghai Science and Technology Committee (STCSM) under Grant No.~19ZR1403000; 
the Ministry of Education, Youth and Sports of the Czech
Republic under Contract No.~LTT17020;
Horizon 2020 ERC Advanced Grant No.~884719 and ERC Starting Grant No.~947006 ``InterLeptons'' (European Union);
the Carl Zeiss Foundation, the Deutsche Forschungsgemeinschaft, the
Excellence Cluster Universe, and the VolkswagenStiftung;
the Department of Atomic Energy (Project Identification No. RTI 4002) and the Department of Science and Technology of India; 
the Istituto Nazionale di Fisica Nucleare of Italy; 
National Research Foundation (NRF) of Korea Grant
Nos.~2016R1\-D1A1B\-01010135, 2016R1\-D1A1B\-02012900, 2018R1\-A2B\-3003643,
2018R1\-A6A1A\-06024970, 2019K1\-A3A7A\-09033840,
2019R1\-I1A3A\-01058933, 2021R1\-A6A1A\-03043957,
2021R1\-F1A\-1060423, 2021R1\-F1A\-1064008;
Radiation Science Research Institute, Foreign Large-size Research Facility Application Supporting project, the Global Science Experimental Data Hub Center of the Korea Institute of Science and Technology Information and KREONET/GLORIAD;
the Polish Ministry of Science and Higher Education and 
the National Science Center;
the Ministry of Science and Higher Education of the Russian Federation, Agreement 14.W03.31.0026, 
and the HSE University Basic Research Program, Moscow; 
University of Tabuk research grants
S-1440-0321, S-0256-1438, and S-0280-1439 (Saudi Arabia);
the Slovenian Research Agency Grant Nos. J1-9124 and P1-0135;
Ikerbasque, Basque Foundation for Science, Spain;
the Swiss National Science Foundation; 
the Ministry of Education and the Ministry of Science and Technology of Taiwan;
and the United States Department of Energy and the National Science Foundation.

\bibliographystyle{apsrev4-2}
%

\end{document}